\documentclass[usenatbib]{mn2e}
\usepackage{times}
\usepackage{epsfig}
\usepackage{threeparttable}
\newcommand{\kms}{~km~s$^{-1}$}

\def \etal      {{\it et al.\ }}
\def \ie        {{\it i.e.,\ }}
\def \eg        {{\it e.g.,\ }}

\def \Mb        {{\rm\ M_{B}}}

\def \Mk        {{\rm\ M_{K}}}
\def \Mj        {{\rm\ M_{J}}}

\begin{document}
\title[Bimodal Infrared luminosity functions in galaxy groups]
{The Group Evolution Multiwavelength Study (GEMS): The near-infrared 
luminosity function
of nearby galaxy groups}
\author[Trevor A. Miles et al.]{%
       \parbox[t]{\textwidth}{Trevor A. Miles\thanks{E-mail:
tm@star.sr.bham.ac.uk}, 
Somak Raychaudhury \& Paul A. Russell}
\vspace*{6pt} \\
School of Physics and Astronomy, University of Birmingham, Edgbaston,
Birmingham B15 2TT, UK}
\date{
MNRAS submitted - 2006 January; accepted - 2006 September
}
\pagerange{\pageref{firstpage}--\pageref{lastpage}}
\pubyear{2006}
\label{firstpage}
\maketitle
\begin{abstract}
We present $J$ and $K$-band luminosity functions (LF) for the Group
Evolution Multiwavelength Study (GEMS) sample of 60 nearby groups of
galaxies, with photometry from the 2MASS survey.  We find that, as
seen in $B$ and $R$-band photometry of a subsample of these groups in
our earlier work, the LFs of the X-ray dim groups ($L_X\!
<\!10^{41.7}$ erg s$^{-1}$) show a depletion of galaxies of
intermediate luminosity around $\Mk\!=\!-23$, within a radius
$0.3\,R_{500}$ from the centres of these groups.  This feature is not
seen in the X-ray brighter groups, nor in either kind of group when
the LFs are determined all the way out to $R_{500}$.  We conclude that
an enhanced level of star formation is not responsible for the this
feature. From the faint end of the LFs, we find support for the
under-abundance of low surface brightness dwarfs in the 2MASS survey.
We find that for all kinds of groups, the modelling of the luminosity
function, with universal forms for the LFs of galaxies of different
morphological types, fails when simultaneously required to fit the $B$
and $K$-band LFs. This means that the dip-like features seen in LFs
are not merely due to the varying proportions of galaxies of different
morphological types among the X-ray dim and bright groups.  We argue
that this support our hypothesis that this feature is due to the
enhanced merging of intermediate-mass galaxies in the dynamically
sluggish environment of X-ray dim groups.

\end{abstract}


\section{Introduction}
Luminosity functions of galaxies are an essential ingredient in the
study of galaxy formation. Numerical models involving merging, cooling
and feedback \citep[\eg][]{somerville99, kauff99, benson03} produce
definite predictions for the luminosity function in various
environments, which can now be compared with photometric observations
of wide-field and deep samples of galaxies of the field and in highly
clustered regions.

The galaxian luminosity function (LF) is usually modelled as the
Schechter function, which drops sharply at bright magnitudes, but
rises at the faint end as a power-law of slope $\alpha$.  In cold dark
matter (CDM) dominated models; dwarf galaxy haloes collapsing at
$z\!>\!3$ have very efficient cooling, resulting in a steep slope
$\alpha \!\approx\! -2$ at the faint end of the mass function
\citep[\eg][]{evrard89,wf91,kwg93}. However, in practice, 
observed galaxy samples over large samples usually yield
$\alpha\!\approx\! -1$
\citep[\eg][APM-Stromlo, $b_J$-band]
{apmstromlo95}. Steeper faint-end slopes have been 
found 
\cite[\eg][]{tt02,2df-lf,trentham05}, 
and there is some indication that this is dependent on environment,
and on the photometric filter used \citep[\eg][]{blanton05}. Ways of
reconciling LFs of the luminous component with the predicted mass
function of dark haloes have variously invoked effects of the physics
of cooling and ionization \citep[\eg][]{chiu01,somerville02}.

Recently, with the help of deep photometry of the nearby Universe, the
faint end of the LF has been explored in various surveys, revealing
features that makes it difficult to fit a single Schechter
function to the entire data \citep[\eg][]{ttm06,blanton05}. 
Indeed, many recent studies suggest that
the LF of brighter galaxies ($M_B\!<\!-18.5$) should be modelled as a
Gaussian, while that of the fainter ones be represented as a
Schechter function \citep[\eg][]{jerjen97,mtt05}. This approach seeks
to explain the peaks and dips in the LF as due to a varying mix of
galaxies of different morphological types in different environments,
and highlights the connection between the evolution of galaxies and
their local environment.

A good example is the luminosity function of galaxies in 
Hickson compact groups, where a prominent deficit
(``dip'') was found at intermediate luminosities \citep{huns98},
showing that a single Schechter function would be inadequate 
to describe these LFs.
Indeed, even in the Coma cluster, various studies
\citep[\eg][]{tg93,lobo97,tully05} have found that 
optical LF of galaxies in the cluster have features not described by a
Schechter function.

In a previous paper \citep{miles04}, we presented optical
($B$ and $R$-band) LFs of a sample of 25 nearby groups of
 galaxies, as part of the Group Evolution Multiwavelength Study 
\citep[][GEMS]{forbes06,osmond04,khos04,rm06}, 
for which we have X-ray luminosities 
(some have upper limits) from ROSAT PSPC observations.  To investigate
the overall shape of the LF, we stacked together the LFs of several
groups, with galaxies chosen from within $0.3\,R_{500}$ from the
centre of each group.  We found that the LF of the X-ray dim groups
($L_X \!<\! 10^{41.7}$ erg~s$^{-1}$) were significantly different from
those of the X-ray brighter groups, showing a prominent dip, similar
to Coma, at around $\Mb \!=\!-18$. We interpreted this deficiency of
intermediate-luminosity galaxies as evidence of rapid evolution
through merging, since X-ray dim groups have lower velocity
dispersion, which would encourage tidal interaction and merging by
means of enhanced dynamical friction.

There have been alternative explanations suggested for the various
features seen in LFs. Some have argued 
\citep[\eg][]{fergsan91,jerjen97} that the total luminosity
function of a group or cluster has features due to a varying
morphological mix across various systems.  Others
\citep[\eg][]{biv95} have suggested that the apparent dip in the optical LF 
might actually be an enhancement of the bright end of the LF, due to excess
star formation caused by tidal effects in the inner
regions of the cluster or group.

We address these issues in this paper. To investigate whether the
shape of the optical LF is affected by variation in the star formation
properties of galaxies, we examine the near-infrared LFs of a large
well-defined sample of groups of galaxies, namely the 60 groups of the
GEMS study. The JHK
photometry from the 2MASS survey is likely to trace evolved stellar
populations and hence the total stellar content of galaxies, 
and thus can be better compared with galaxy formation models. 
It also allows us to
avoid the effects of extinction that plague optical studies.

In the following section, we analyse 2MASS photometry of galaxies
belonging to the 60 GEMS
groups, and, by splitting them into two categories based on
their X-ray luminosity, compare their $K$-band LFs with our earlier
$B$-band LFs for a subset of 25 groups \citep{miles04}.  In \S3, we
examine how the luminosity functions of X-ray bright and dim groups
change with projected radial distance from group centre. We address
the issue of the shape of the LF depending on the mix of various
morphological types in \S4, where we compute model luminosity
functions resulting from the summation of the LFs of individual galaxy
types. In the final section, we discuss the implications in the
context of the evolution of galaxies in groups.  We have used $H_0=70$
\kms Mpc$^{-1}$; $\Omega_0\equiv\Omega_M=1$ throughout.

\section{Near-Infrared Luminosity functions of groups}

Our sample of 60 nearby groups of galaxies, known as the Group Evolution
Multi-wavelength Study (GEMS), is described in detail in
\cite{osmond04} and \cite{forbes06}. 
It was compiled to represent a wide variety of groups over a large
range stages of evolution. First, a master sample of of over 4000
optically identified groups was constructed from the literature, from
which only those that were found to be within the field of a ROSAT
PSPC pointed observation with integration time $>$10~ks, were chosen
as members of the GEMS sample.  A large fraction of the GEMS groups
were found to have been detected in these pointed PSPC observations,
and for the remaining few we have upper limits for their X-ray flux.

\begin{figure}
\epsfig{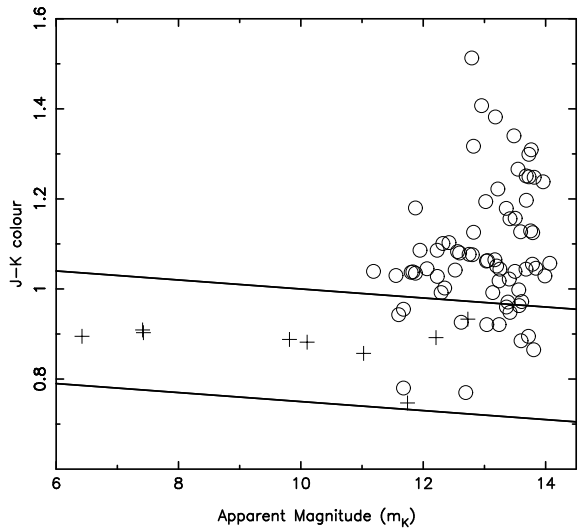}
\caption{An illustration of the colour-magnitude selection, showing the 
NGC~4636 group, which has the highest number of known galaxy redshifts
in the GEMS sample.  Group members, ascertained on the basis of
measured redshift, are plotted as crosses, and non-group galaxies as
circles.  The range of the expected red sequence, as described in the
text, is plotted as a pair of lines.  The non-group galaxies, within
the width of the red sequence down to our magnitude limit
($K\!=\!13.5$), will be removed by our background subtraction method
shown Fig.~\ref{fig:fieldlf}}
\label{fig:cm4636}
\end{figure}

\begin{figure}
\epsfig{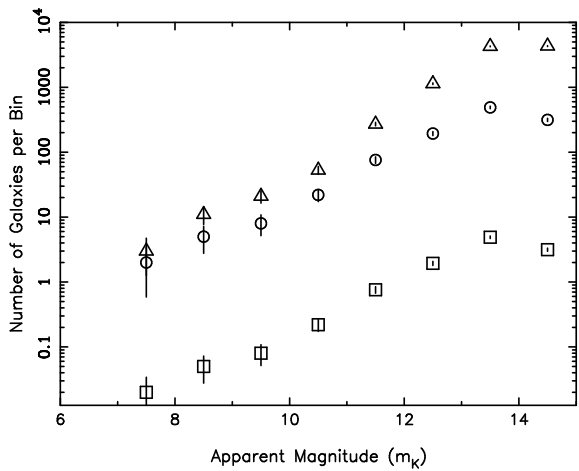}
\caption{Estimating the background: for each group, 
we select 100 random fields (of the same area as a $R_{500}$ circle for
that group) around it, and compute the total number of galaxies in
each $K$-band apparent magnitude bin.  Here, triangles represent such
a plot for the NGC~5044 group.  The circles represent those galaxies
that survive the colour-magnitude selection criterion applied from
that group.  Finally, the squares represent the circles
divided by one hundred, this being the ``background'' histogram 
to be subtracted when constructing the LF for this group. The trail-off
at the faint end shows the incompleteness of the 2MASS catalogue
fainter than $\Mk$=13.75 \citep{jar2000}.}
\label{fig:fieldlf}
\end{figure}

\subsection{Galaxy Selection and background subtraction}

The 2MASS All-Sky Data Release catalogue contains more than 1.6
million extended sources, with a limiting magnitude of $\Mk$=13.75
\citep{jar2000}. For each GEMS group, we selected
all objects, belonging to the 2MASS All-Sky Extended Source Catalog
(2MASX), within a circle of radius $R_{500}$ (radius out to which the
mass overdensity is 500 times the mean density of the Universe) of its
centre. The adopted group centres, values of $R_{500}$ and distances
to these groups can be found in \citet{osmond04}.

The $J\!-\!K$ colours were calculated from the isophotal magnitudes
($K_s$= 20 mag arcsec$^{-2}$).
Galaxies were selected as being likely group members from the red sequences of
the colour-magnitude diagram, involving the $(J-K)$ colour
and the $K_s$ magnitude 
The width of the red sequence
was based on the Coma Cluster $U\!-\!V$
colour-magnitude relation of \cite{tcb01}, its
3$\sigma$ width being
converted to a $J\!-\! K$ width based on \cite{jar2000}. 

Fig.~\ref{fig:cm4636} illustrates this for the case of the NGC~4636
group, which has the highest number of known galaxy redshifts in the
GEMS sample. It has 80 measured redshifts within $R_{500}$
(most GEMS groups have under 10 measured redshifts 
within this radius and 21 groups have
under 5 redshifts).  The colour-magnitude diagram shows group galaxies
as crosses, and non-group galaxies as circles.  The non-group galaxies
that are within the $3-\sigma$ are all faint and are then removed by
our background subtraction method.  out to the magnitude limit used
here ($K\!<\!13.5$).

Given that the number of galaxies in each group is small, the
luminosity functions may be affected significantly by the method of
background subtraction. For this reason, for each group, we took a
large sample of field galaxies from 100 randomly-selected,
independent, regions on the sky, each of radius equal to the
value of $R_{500}$ for the group. We then calculated the mean number of 
galaxies occupying each magnitude bin and subtracted them from the target 
group. The whole process was then repeated for every group in the GEMS sample. 

Fig.~\ref{fig:fieldlf} shows one such example, where the raw number of
galaxies found in these field regions is shown as triangles, binned in
apparent $K_s$ magnitude. Open circles show all field galaxies from
these 100 regions satisfying the C-M selection criterion for the
group. The open squares show those galaxies surviving the C-M cut
divided by 100, representing the average number of local background
and foreground galaxies expected in each bin of apparent magnitude.
In this way we aimed to account for cosmic variance and to have
compiled as 
unbiased a selection of field galaxies as possible.

\section{Luminosity functions of X-ray bright and dim groups}

Since the number of member galaxies in each individual group is small,
the galaxian luminosity function (LF) of the groups was evaluated by
co-adding galaxies of several groups in equally spaced bins of
absolute luminosity.

An apparent magnitude cut-off was adopted at $K\!=\!13.5$, since
Fig.~\ref{fig:fieldlf} shows that the samples of 2MASS galaxies, in
100 randomly-selected field regions, is reasonably complete to
$K\!\sim$13.75 \citep[cf.][]{jar2000}.  The resulting limit in
absolute magnitude varies between groups-- the distance moduli of the
60 groups range between 30.1 and 35.5, with the median of 32.5.
Hence, co-added LFs were evaluated by dividing the number of co-added
galaxies in each bin of absolute magnitude by the number of
contributory groups to that bin.  When plotting the group LFs, we
chose a conservative absolute magnitude limit of $M_K<-20$ so that the
faintest bin contains over half of groups within the 2MASS
completeness limit of $K\!<\! 13.5$.

Various optical and X-ray properties of GEMS groups may be found in
\cite{osmond04} and \cite{forbes06}.  
The groups in our sample represent very diverse systems in terms of
their content and physical properties. We divide our sample of
galaxies into two sub-classes according to the X-ray luminosity of
their parent groups, which is related to the mass and velocity
dispersion of the latter.

As in \cite{miles04}, 
we use the X-ray luminosities measured by
\citet{osmond04}, who used ROSAT PSPC observations in the 0.5--2 keV
range, fitting $\beta$-profiles after point source removal,
extrapolated to estimate the bolometric X-ray luminosity.  We
characterised the parent groups as X-ray bright if their bolometric
X-ray luminosity is more than the median of the sample,
$L_X=10^{41.7}$ erg~s$^{-1}$, and X-ray dim if less. This X-ray
luminosity refers to that of the group plus of any central galaxy that
might exist.

\subsection{Stacked Near-Infrared Luminosity Functions within 0.3$R_{500}$}

\begin{figure*}
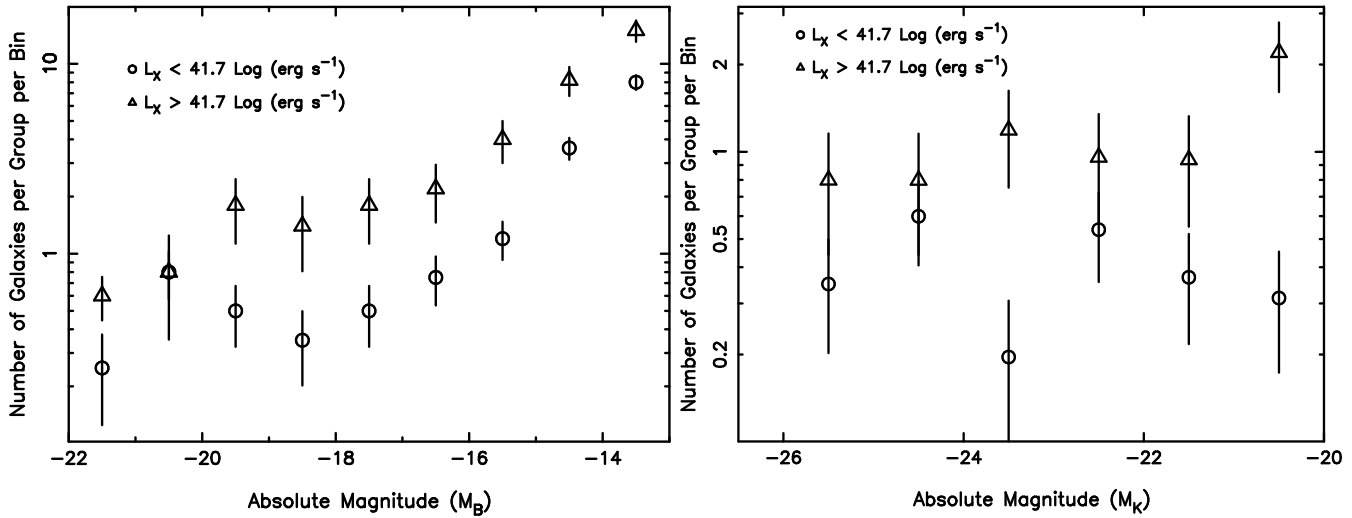

\centerline{
\epsfig{file=TMfig3.ps,angle=-90,width=0.5\hsize}
\epsfig{file=TMfig3b.ps,angle=-90,width=0.5\hsize}
}
\caption{The mean 
differential Luminosity Functions of 25 GEMS Groups of galaxies, in the 
$B$-band (left) and $K$-band (right), each within
a radius of 0.3$R_{500}$ from their respective centres,
stacked together to form a composite LF for the respective sub-classes.
These groups are the same as those described in Miles \etal\ (2004).
Here X-ray bright groups ($L_X>10^{41.7}$ erg~s$^{-1}$,
triangles) and X-ray dim groups ($L_X<10^{41.7}$ erg~s$^{-1}$,
circles.  It is clear that the LFs of the dim groups show ``dips'' in
the LFs between $-19<M_B<-17$ and $-24<M_K<-23$.
\label{fig:diffcomp}}
\end{figure*}

The differential galaxian Luminosity Functions of the 25 GEMS Groups
that made up our earlier optical sample \citep{miles04} above are
shown in Fig.~\ref{fig:diffcomp} in $B$ and $K$-bands, out to
$0.3\,R_{500}$, for direct comparison between the optical and near-IR
LFs.  As in the optical LFs ($B$, as shown here, and in $R$) analysed in 
\citet{miles04}, a clear difference
between the LFs of the 
X-ray bright groups and X-ray dim groups is evident,
with a depletion in the number of galaxies with magnitudes
between $-24\!<\!{\Mk}\!<\!-23$ and $-23\!<\!{\Mj}\!<\!-22$ for the
latter category of groups.  The possibility that these dips are due to
some of the galaxies being systematically missed can be ruled out. As
is apparent from Fig.~\ref{fig:fieldlf}, our 2MASS sample is
complete to about four magnitudes fainter than the position of the dip
in the luminosity function.

The previous pair of figures represented the combined LFs in the two
categories out to $0.3\,R_{500}$ from the centres of only the 25 groups
covered in the \citet{miles04} study, so that the $K$-band LF could be
directly compared with the optical results.  In the optical
study, however,
the number of groups studied and the area of sky covered in the
case of each group were limited by the availability of observing time,
and the size of the wide-field CCD arrays we had been using. 

\begin{figure*}
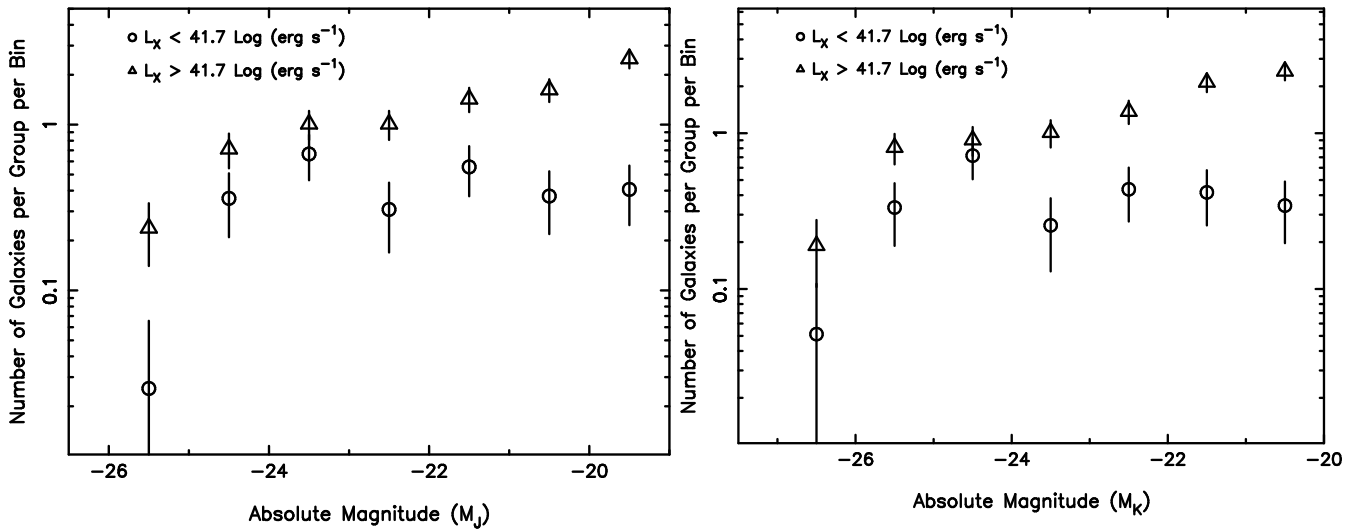

\centerline{
\epsfig{file=TMfig4.ps,angle=-90,width=0.5\hsize}
\epsfig{file=TMfig4b.ps,angle=-90,width=0.5\hsize}
}
\caption{
The near-infrared mean differential 
Luminosity Functions of the entire GEMS sample- all 60
Groups of galaxies, in the 
$J$-band (left) and $K$-band (right), each within
a radius of 0.3$R_{500}$ from their respective centres,
stacked together to form a composite LF for the respective sub-classes.
As in Fig.~\ref{fig:diffcomp},
X-ray bright groups ($L_X>10^{41.7}$ erg~s$^{-1}$ are plotted as 
triangles, and X-ray dim groups ($L_X<10^{41.7}$ erg~s$^{-1}$) as
circles.  It is clear that the LFs of the dim groups show ``dips'' in
the LFs between $-23<M_J<-22$ and $-24<M_K<-23$.
\label{fig:diffcomp-whole}
}
\end{figure*}

In this paper, our near-IR magnitudes come from the 2MASS survey, so
we can deal with the whole GEMS sample of 60 groups, as well as go out
in radius further than the previous study. We do the latter in the
next section, but here in Fig.~\ref{fig:diffcomp-whole} we plot the
$J$ and $K$-band differential Luminosity Functions of the entire GEMS
sample- all 60 Groups of galaxies, including those plotted in
Fig.~\ref{fig:diffcomp}, each within a radius of 0.3$R_{500}$ from
their respective centres, stacked together to form a composite LF for
the respective sub-classes.  There are 39 X-ray dim and 21 X-ray
bright groups in this sample, and the total number of
galaxies in either category is similar.

As in Fig.~\ref{fig:diffcomp}, it is
clear that the LFs of the dim groups show ``dips'' in the LFs between
$-23<M_J<-22$ and $-24<M_K<-23$, thus showing that the GEMS sub-sample
chosen in the \citet{miles04} work was representative, and that the
near-IR LFs are similar, irrespective of the filter used.

One possible reason for finding a dip feature fainter than the bright
end of the LF of groups is that it could result from the fact that is
most poor groups, there is one dominant galaxy, which would account
for an enhancement of galaxies towards the bright end of the LF. We
examine whether this is the case of the GEMS LFs, by plotting in
Fig.~\label{fig:nobgg}, the $K$-band differential Luminosity Function
of all 60 groups (same as in Fig.~\ref{fig:diffcomp-whole}b), but
without the brightest group galaxies, plotted on the same scale as
Fig.~\ref{fig:diffcomp-whole}b) to allow direct comparison. As
expected, no galaxies remain in the brightest bin, and there are fewer
in the second brightest, but the dip feature remains, showing that the
brighest group galaxies do not account for the appearance of this
feature.  The LFs without the central bightest galaxy are similar in
the other optical and near-IR filters.

\subsection{Luminosity Functions as a Function of Radius}

\begin{figure*}
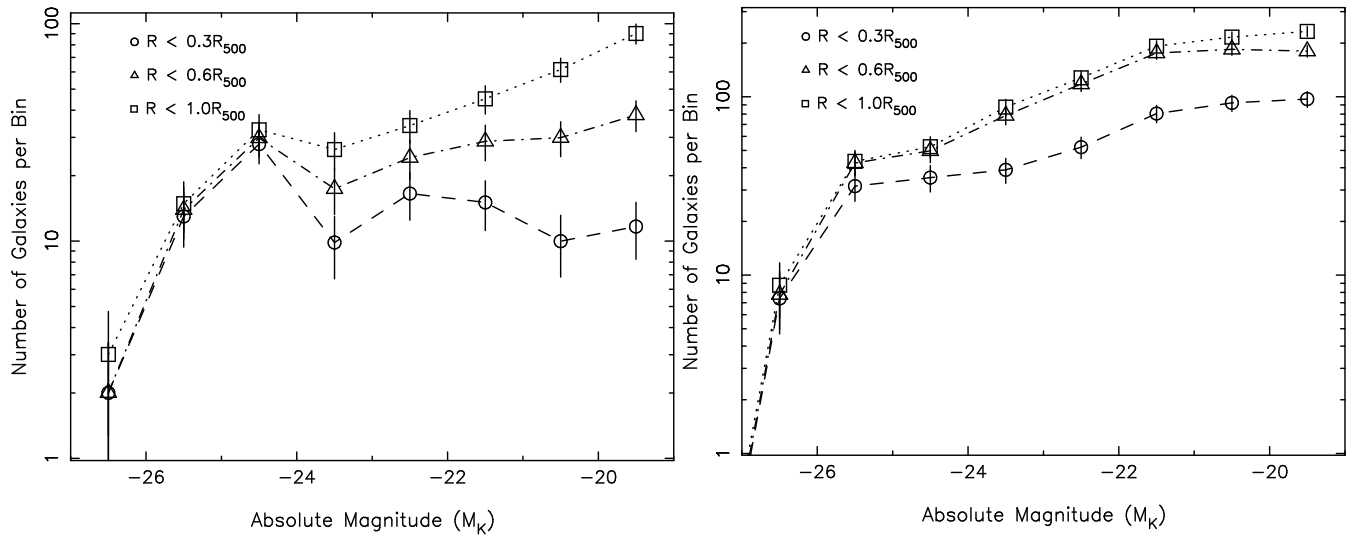

\centerline{
\epsfig{file=TMfig6.ps,angle=-90,width=0.5\hsize}
\epsfig{file=TMfig6b.ps,angle=-90,width=0.5\hsize}
}
\caption{
$K$-band composite differential luminosity function of (Left:) all 39
X-ray dim groups ($L_X<10^{41.7}$ erg~s$^{-1}$), interior to a
fraction of the projected group radius, and (Right:) all 21 high X-ray
luminosity groups ($L_X>10^{41.7}$ erg~s$^{-1}$). The three plots in
each case go out to 0.3, 0.6 and 1.0 times $R_{500}$ respectively.
The intermediate luminosity dip feature is more prominent in the inner
regions of the X-ray dim groups. The LF of the X-ray bright groups remains
similar in shape at all radii.}
\label{fig:lowlfrad}
\end{figure*}

Differential $K$-band Luminosity Functions of all X-ray dim and
X-ray bright GEMS groups are shown in Fig.~\ref{fig:lowlfrad}.  The
groups in each category are stacked together in radial bins upto
0.3$\,R_{500}$, upto 0.6$\,R_{500}$ and upto 1.0$\,R_{500}$.
It is clear that the LF of the X-ray dim groups show a deficiency in
the LF between $-24\! <\! M_K \!<\! -23$ within 0.3$R_{500}$, together
with a flattening in the fainter galaxies at magnitudes $M_K>-22$.

\begin{figure*}
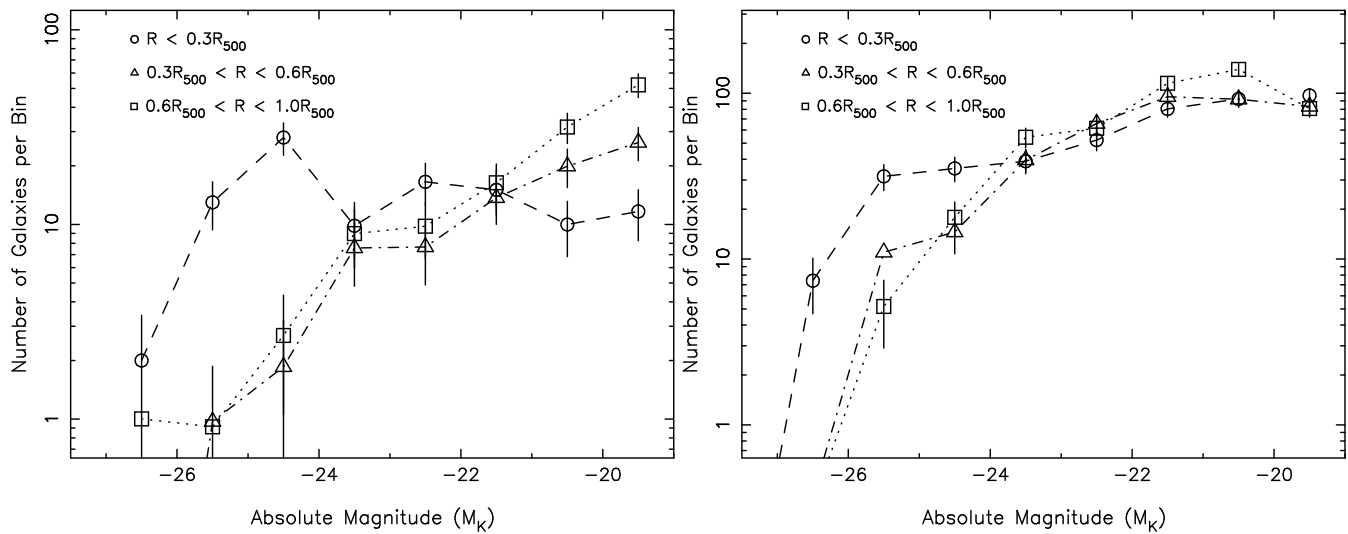

\centerline{
\epsfig{file=TMfig7.ps,angle=-90,width=0.5\hsize}
\epsfig{file=TMfig7b.ps,angle=-90,width=0.5\hsize}
}
\caption{As in the previous figure, but in differential radial bins.
(Left:) The $K$-band luminosity function of all X-ray dim groups
($L_X\!<\!10^{41.7}$ erg~s$^{-1}$) and (Right:)
all  X-ray bright groups
($L_X\!>\!10^{41.7}$ erg~s$^{-1}$), divided into annuli of
projected radius. The inner 0.3$R_{500}$ of the X-ray dim groups shows
a luminosity function of a significantly different shape, in comparison
with both the high X-ray luminosity groups at all radii and also with
the outer regions of the X-ray dim groups themselves.}
\label{fig:difrad}
\end{figure*}

Fig.~\ref{fig:difrad} shows in detail how dividing the
sample into radial annuli, rather than the integrated average
out to a certain radius as shown in
Fig.~\ref{fig:lowlfrad}, illustrates explicitly the difference in LF
shape in the central regions of X-ray dim groups.  The dip between
$-24<M_K<-23$ seen in the LF in the central regions of the groups gradually
disappears as the 
LF is averaged out to larger radii,
approaching $R_{500}$.

Furthermore, when averaged out to large distances ($\sim\,R_{500}$)
from the centres of the groups, X-ray bright and X-ray dim groups have
similar LF shapes (see Fig.~\ref{fig:lowlfrad}). Taking all galaxies
out to $R_{500}$, the K-S test shows that it is not possible to
distinguish between the two samples at the 92\% confidence level.

\begin{figure}
\epsfig{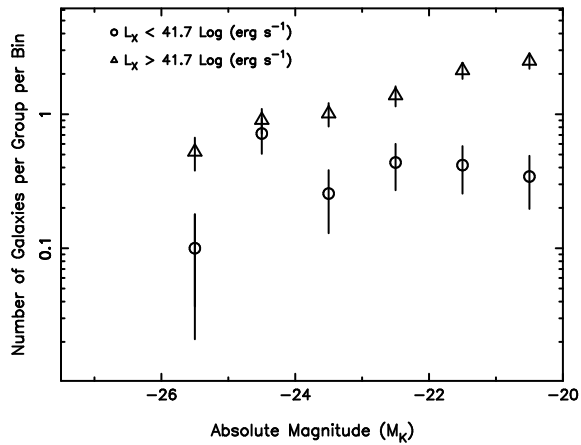}
\caption{The $K$-band differential Luminosity Function of the GEMS sample,
 without the brightest galaxy in each group, 
plotted on the same scale as Fig.~\ref{fig:diffcomp-whole}b
 to allow direct comparison.
The brightest two bins are diminished, whilst the dipping feature 
found in Fig. \ref{fig:diffcomp-whole} remains.}
\label{fig:nobgg}
\end{figure}

\subsection{The slope at the faint end of the LF}

Another feature of the LF to note in Figs.~\ref{fig:diffcomp} and
\ref{fig:diffcomp-whole} is the relative flattening of the LF in the
$K$-band  at the faint end compared to the $B$-band, particularly in the 
in the cores of the X-ray dim groups
($<\!0.3\,R_{500}$). One possibility is that it could result from
enhanced recent star formation
activity being revealed in the $B$-band and not in the $JHK$ bands. 
 Tidal
interactions in the densest regions of these dynamically young groups
may be responsible for preferentially inducing star formation in the
less massive galaxies.

\begin{figure}
\centerline{
\epsfig{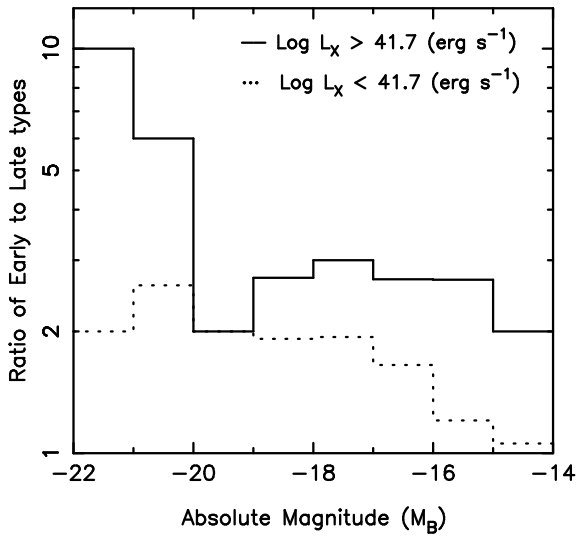}
}
\caption{The ratio of early to late type galaxies (where morphology
is assigned by visual inspection from optical CCD images), 
divided into X-ray bright
($L_X>10^{41.7}$ erg~s$^{-1}$) and  X-ray dim 
($L_X<10^{41.7}$ erg~s$^{-1}$) groups. The relative abundance
dominance of late-type dwarf
galaxies in the X-ray dim groups may
account for the flattening of the faint end of their
near-IR luminosity function, if low surface brightness
objects are being systematically missed in the 2MASS survey.}
\label{fig:ratio}
\end{figure}

Another reason for the difference in faint end slope between $B$ and
$K$ band data for X-ray dim groups could be the bias against low
surface brightness galaxies in the 2MASS catalogue, as discussed
by the original study of galaxies in the survey
\citep{bell03}. We test this by examining the morphology of
all member galaxies in the $B$ image individually by eye, and
classifying them as early or late-type. Fig~\ref{fig:ratio} shows the
ratio of early to late type galaxies in absolute $B$ magnitude bins,
for the X-ray bright groups (solid line) and X-ray dim groups (dotted
line). 

This shows that at the faint end, the fraction of early-type galaxies,
as identified from optical images, is significantly higher in the
X-ray bright groups.  The faint end of the LF of X-ray faint groups
has a higher dominance of late-type galaxies than in X-ray bright
groups.  If low surface brightness late-type galaxies are
systematically missed in the 2MASS survey, this would account for the
flattening of the LF at the faint end, compared to that in the
$B$-band,


\section{The Morphological content of Groups}

It is important to compare our results with previous studies
utilising the 2MASS catalogue.
\cite{bal01} use the second incremental release data
from 2MASS to build a composite LF for clusters, based on 274 cluster
galaxies with redshift measurements from the Las Campanas Redshift
Survey. Despite the large scatter due to the small number of galaxies
in their sample, they find best fit values for the Schechter
parameters $M_{K*}\!=\! -23.8\pm 0.4$ and $\alpha \!=\! -1.30\pm
0.43$ in clusters, $M_{K*}\!=\! -23.5\pm 0.1$ and $\alpha \!=\!
-1.14\pm 0.26$ in groups and $M_{K*}\!=\! -23.4\pm 0.1$ and $\alpha
\!=\! -1.12\pm 0.21$ in the field.  On the other hand, \cite{lin04}
use the full 2MASS data release to produce $K$-band LFs of galaxies
within 93 galaxy clusters and groups, including the three GEMS groups,
NGC~2563, NGC~6338 and HCG~62. They found difficulties fitting a
single Schechter function to their data and resorted to the removal of
the brightest galaxies from their fitting proceedure to produce
$M_{K*}\!=\! -24.02\pm 0.02$ and $\alpha \!=\! -0.84\pm 0.02$. i.e.,
they obtained a higher $M_{K*}$ than \cite{bal01} and a flatter
$\alpha$ parameter.

\subsection{LFs of various morphological types}

Solutions to the problem of the inadequacy of a single Schechter
function fit have been found when meeting dipping features in the LF
of a number of structures ranging in mass from the Coma
\citep{ap00}, Virgo and Fornax clusters \citep{fergsan91} 
to Hickson groups \citep{huns98} and the Leo I group
\citep{flint03}. 

Attempts have been made to explain these features as the consequence
of the observed LF being just the superposition of a number of
underlying LFs, each describing that of a different morphologival type
of galaxies, and the variation in the LF would then just be due to the
variation in the relative abundances of the different contributing
galaxy types. 

To test this hypothesis, we perform the following test. We start from
the the $B$-band LF of both X-ray bright and X-ray dim groups, within
a radius of 0.3$R_{500}$, as obtained in \cite{miles04}. We treat them
as a superposition of five different kinds of LFs, each for a
different morphological type (Ellipticals, S0s, Spirals, Dwarf
Irregulars and Dwarf Ellipticals). We then correct each component LF
for the $B\!-\!K$ colour appropriate for that morphological type and
ask whether the resultant $K$-band LF resembles the observed one.  If
the variation of the observed LFs are solely due to a varying fraction
of the different morphological types, the predicted LFs found this way
should match the observed ones.

The type-specific LFs are not neccessarily of the Schechter form but
may take Gaussian shapes \citep{jerjen97,jerjen-lf}.  In the case of
elliptical, S0 and spiral galaxies, we use Gaussian LFs, and for the
dwarf early and late types we use Schechter functions, following the
\cite{jerjen-lf} prescription (see Table~1).  

\begin{table}
\centering
\caption{Analytical functions and fixed parameters for the 
adopted type-specific Luminosity Functions (following Jerjen 2001).
\label{table1}
}
\begin{tabular}{lrrcr}
\hline
\hline Galaxy Type & Function & Parameter 1 & Parameter 2 \\ 
\hline

 Elliptical&Gaussian&$\overline{M_{B}}$ = -18.3 &$\sigma_{(M<\overline{M_{B}})}$ = 2.2 \\ 
    &     & & $\sigma_{(M>\overline{M_{B}})}$ = 1.3 \\
 S0&Gaussian&$\overline{M_{B}}$ = -18.9&$\sigma$ = 1.1 \\ 
 Spiral&Gaussian&$\overline{M_{B}}$ = -18.3&$\sigma$ = 1.4 \\ 
 dIrr&Schechter&$M_{B}^{*}$ = -16.2&$\alpha$ = -1.0 \\ 
 dE&Schechter&$M_{B}^{*}$ = -17.8&$\alpha$ = -1.4 \\ 
 \hline

\end{tabular} 
\end{table}

\begin{figure*}
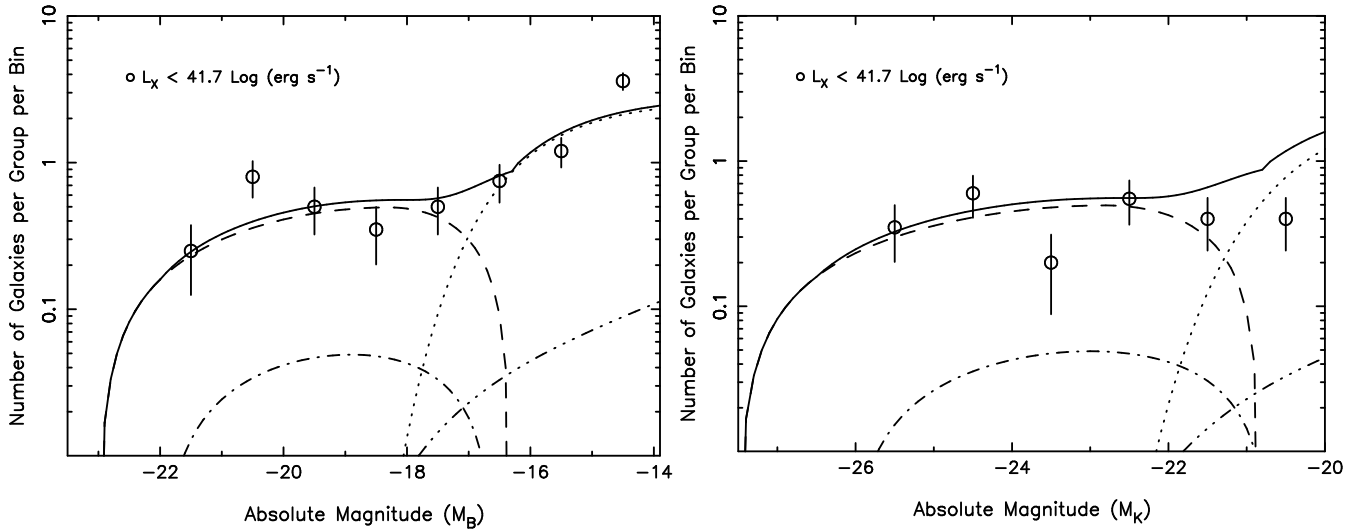

\centerline{
\epsfig{file=TMfig9.ps,angle=-90,width=0.5\hsize}
\epsfig{file=TMfig9b.ps,angle=-90,width=0.5\hsize}
}
\caption{ {\bf [Left]} The best fit five-component 
LFs, one for each morphological type
(Ellipticals, S0s, Spirals, Dwarf Irregulars and Dwarf Ellipticals),
following the \citet{jerjen-lf} functional forms (see Table~1), that add
up to form the best model $B$-band LF. Also plotted are the observed
data, averaged for 20 X-ray dim groups ($L_X<10^{41.7}$ erg~s$^{-1}$)
out to 0.3$R_{500}$. Here the dashed line = Ellipticals; dash-dot = S0s; 
dot-dot-dot-dash = dEs; dotted line = dIrrs \&
continuous line = sum of all components). According to these fits, 
spirals make a negligible contribution to the morphological makeup, contrary
to observations  
{\bf [Right]} The predicted 
$K$-band LFs of the five components, using the same line styles,
together with the observed LF from the 2MASS analysis in this paper.
\label{fig:jerlow}}
\end{figure*}

\begin{figure*}
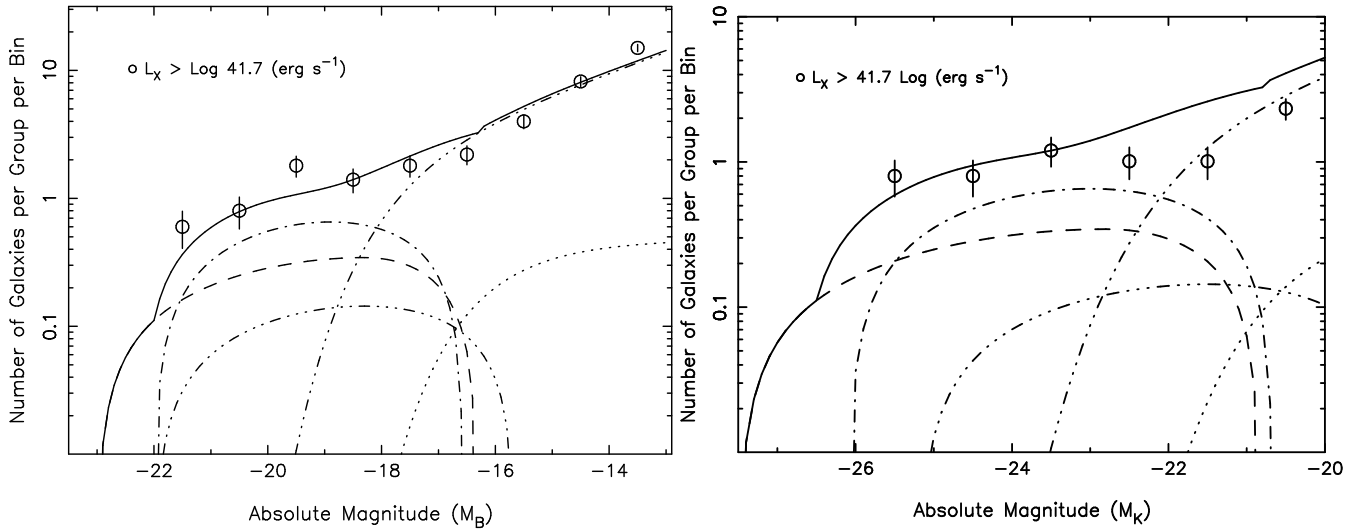

\centerline{
\epsfig{file=TMfig10.ps,angle=-90,width=0.5\hsize}
\epsfig{file=TMfig10b.ps,angle=-90,width=0.5\hsize}
}
\caption{{\bf [Left]}
$B$-band LFs, similar to Fig.~\ref{fig:jerlow}, but for the
X-ray bright groups 
($L_X>10^{41.7}$ erg~s$^{-1}$), 
in our optical sample, out to 0.3$R_{500}$. 
{\bf [Right]} Corresponding predicted 
$K$-band LFs of the individual components, 
together with the observed LF from the 2MASS analysis in this paper.
The line style used for the
various types is the same as before. In this case,
 Spirals (= dash-dot-dot-dot, 
centred on $M_B$ = -18.3) feature in the fitting process.
\label{fig:jerhigh}}
\end{figure*}

Markov chain Monte Carlo (MCMC) fitting was used to determine the
best-fit total $B$-band composite LF from the five component LFs, for
both the X-ray faint (Fig.~\ref{fig:jerlow}a) and X-ray bright groups
(Fig.~\ref{fig:jerhigh}b).  The MCMC algorithm \citep[\eg][]{lb02} is
a computationally efficient way to map the likelihood of the fit to
the data in the high-dimensional parameter space of the five LF
components. The MCMC algorithm was used to draw samples with a number density
asymptotically proportional to the probability density representing
the likelihood of the fit. In this way, the shape of the full
posterior probability density (using a uniform prior)
was explored while avoiding excessive processor time in
low-probability regions of the fit-space.

\begin{table}
\centering
\caption{$B\!-\!K$ colours 
for galaxies used 
to convert $B$-band LFs to $K$-band LFs
in Figs.~\ref{fig:jerlow} and \ref{fig:jerhigh} (after Jarrett 2003).
\label{table2}
}
\begin{tabular}{lr}
\hline
\hline Galaxy Type & $B_{\rm Total}$-$K_{\rm Total}$\\ 
\hline
 Elliptical&4.0\\ 
 S0&4.1\\ 
 Spiral&3.2\\ 
 dIrr&4.1\\ 
 dE&4.0\\ 
\end{tabular} 
\end{table}

The MCMC algorithm selected new points in the fit-space by
simultaneously jumping along all the parameters 
representing the five LF components. The size of this
jump was always a random fraction of a characteristic step-size along
each parameter. The chi-squared statistic at this new point was then
evaluated, and the new point accepted or rejected. The process was
then repeated, jumping again from the most recently accepted point in
the fit-space. The number of accepted points was optimised by
adjusting the step-size along each component. Excessively large step
sizes meant that very few new points were accepted, and overly small
step sizes restricted exploration around the fit-space. The algorithm
had clearly determined a best-fit curve from the five components after
$10^7$ iterations.

Fig.~\ref{fig:jerlow}a shows the results of the fitting procedure,
plotted superposed on the observed $B$-band data, for the X-ray dim
groups.
The best-fit component LFs
were corrected by the colour appropriate for the particular morphological type
using the typical $B\!-\!K$ colours for each morphological type, as
given in
\cite{jar2003} and shown in Table~2, and the predicted composite $K$-band
luminosity function is shown in Fig.~\ref{fig:jerlow}b. Superposed on these
are the observed LFs.

It is obvious that this procedure would not work if it is required to
fit both the $B$ and $K$ band LFs simultaneously.  The $K$-band LFs
produced from the $B$-band observed data do not have the same features
as the observed $K$-band LFs.
In order to find a good match to the observed $K$-band LFs,
one expects the $B$-band LF to be dominated almost exclusively by
ellipticals and S0s brighter than $-18$, and the number of spirals
to be insignificant.

For the X-ray bright groups, shown in Fig.~\ref{fig:jerhigh}, early
types dominate the brighter end, and again there are very few spirals
brighter than $-18$, contrary to what is observed.

\subsection{Comparison with visually verified morphological content}

We can go one step further and check the morphological composition of
these groups from our visual inspection of galaxy images in the X-ray
dim groups.

The visual classification used here is the same as that used in
\cite{miles04}, further details may be found therein.  A simple
classification system was adopted whereby each galaxy was classified
as early- or late-type based on inspecting the $B$-band optical CCD
image. The relative content of early and late type galaxies, thus
determined, is shown in Fig.~\ref{fig:morph}, and compared with the
relative abundances predicted by the LF fits in the previous section.

The component LF-fitting process predicts the presence of large
numbers of dwarf irregulars in the $B$-band (Fig.~\ref{fig:jerlow},
dotted line), which seems to be is a reasonable fit to the data
(Fig.~\ref{fig:jerlow}b (dotted line); Fig.~\ref{fig:morph}b, dotted
line).  However, the model fails to fit the dwarf elliptical
population (dot-dot-dot-dashed line), underestimating the observed
frequency in Fig.~\ref{fig:morph}a.  The spiral galaxy population is
missed entirely by the models (Fig.~\ref{fig:morph}b, dash-dot-dot-dot
line).  The predicted early-type population (elliptical and
lenticulars) matches the observed data better than the late-type curve but
also fails to pass through any data point within error bars.

The apparent failure of the component morphology-based LF curves to fit
real data indicates that the galaxy population in groups of galaxies
has a different morphological mix from that of clusters and in the
field. Evolutionary processes are most favoured in groups and may be
responsible for the differences in galaxy populations observed.

\section{Discussion and Conclusions} 

The main problems associated with Luminosity Functions at optical
wavelengths are that they are liable to be affected by dust extinction
as well as by young stellar populations, thus rendering the light less
representative of the underlying matter.
This is far less of a problem in the near-infrared,
where one is more likely to trace the more evolved stellar populations
and hence the actual stellar content of galaxies.  Near-infrared
luminosities are more directly related to stellar mass, constraining
both the history of the star formation and galaxy formation models.  
The fact that we find dip features in the
luminosity functions of X-ray dim groups of galaxies in the
near-infrared regime means that it is unlikely that star formation or
dust extinction is responsible for the observed depletion of intermediate
magnitude galaxies.

Such features have often been explained as the result of the
representation in the total galaxy luminosity function of the
underlying morphological makeup of groups.  Invariant type-specific
LFs have been claimed \citep[\eg][]{jerjen-lf}. Implicit in this
notion is the assumption of passive evolution at all structure scales,
and thus the shape of the LF is merely due to the combination of the
individual LFs of galaxies of different morphological types in varying
proportions.  The deviation from these functional forms may therefore
provide evidence of predominant major mergers affecting this model,
and bringing into question the invariance of the LFs of individual
morphological types.

\begin{figure*}
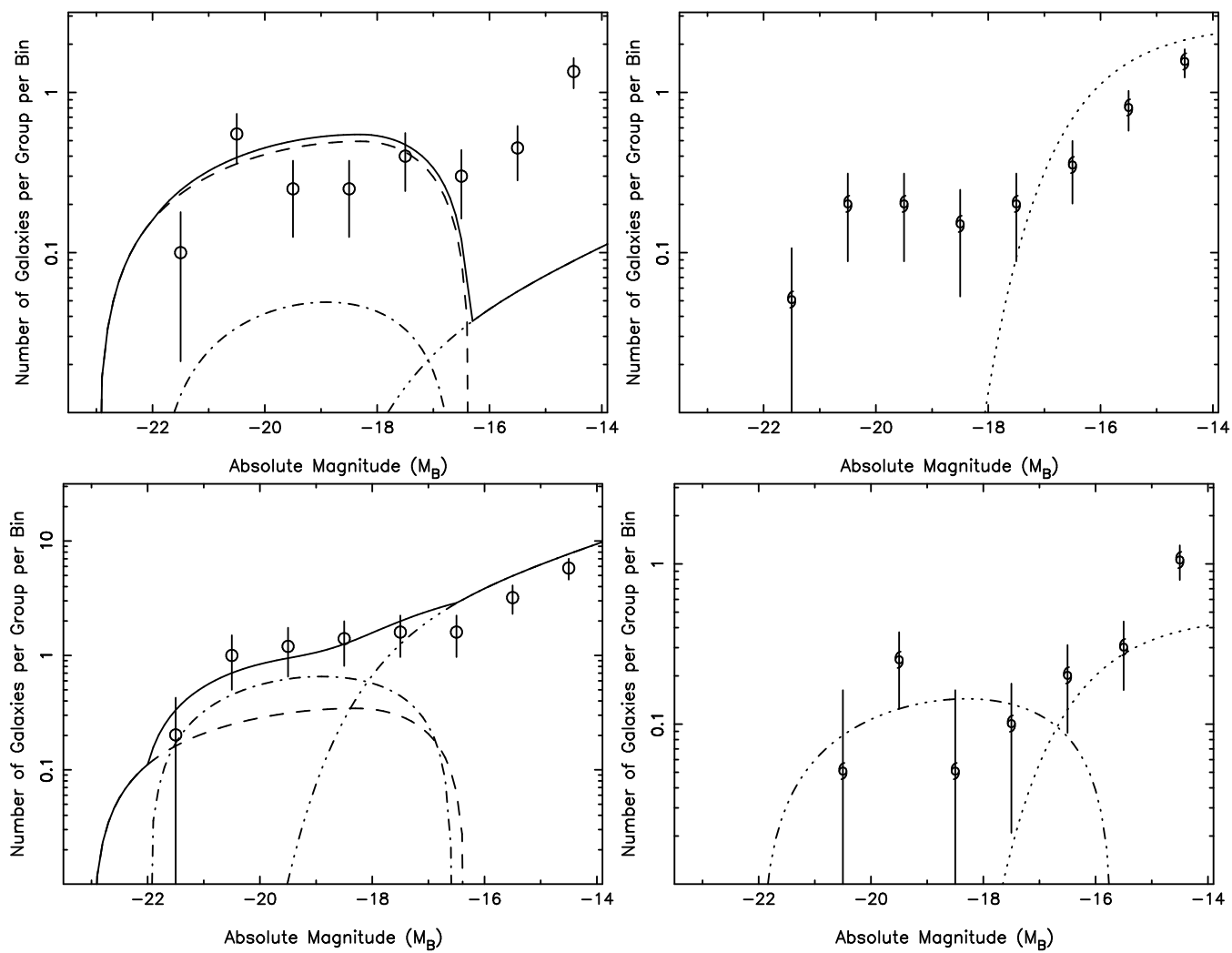

\centerline{
\epsfig{file=TMfig11.ps,angle=-90,width=0.5\hsize}
\epsfig{file=TMfig11b.ps,angle=-90,width=0.5\hsize}}
\centerline{\epsfig{file=TMfig11c.ps,angle=-90,width=0.5\hsize}
\epsfig{file=TMfig11d.ps,angle=-90,width=0.5\hsize}
}
\caption{The four panels show observed $B$-band
luminosity functions for the Early-type galaxies (circles)
and late-type galaxies (spirals).  The top two panels 
are for X-ray dim groups ($L_X\!<\!10^{41.7}$ erg~s$^{-1}$), and
the two bottom panels for X-ray bright groups
($L_X\!>\!10^{41.7}$ erg~s$^{-1}$), both
out to 0.3$R_{500}$.   Shown for comparison are the best fit 
component LFs
from Figs.~\ref{fig:jerlow} and \ref{fig:jerhigh}.
{\bf [Left]} Ellipticals = dashed line; S0s = dash-dot; dEs = dash-dot-dot-dot; continuous line = sum of all components.
{\bf [Right]} Spirals = dash-dot-dot-dot; dIrrs = dotted line.
\newline {\bf [Top Left]} Data points: Early-type galaxy  $B$-band LF for X-ray dim groups. 
Curves: Early-type components fitted to total  X-ray dim group LF. 
\newline {\bf [Top Right]} Data points: Late-type galaxy  $B$-band LF for X-ray dim groups. 
Curves: Late-type components fitted to total X-ray dim group LF. 
\newline {\bf [Bottom Left]} Data points: Early-type galaxy  $B$-band LF for X-ray bright groups.
Curves: Early-type components fitted to total  X-ray bright group LF. 
\newline {\bf [Bottom Right]} Data points: Late-type galaxy  $B$-band LF for X-ray bright groups.
Curves: Late-type components fitted to total  X-ray bright group LF.
  This exercise shows that the
morphological abundances from the best-fit model LFs do not match the
observations.
\label{fig:morph}}
\end{figure*}

Fig.~\ref{fig:morph} shows the observed luminosity function of
galaxies visually classified as early and late types.  The superposed
curves are the results from the earlier fitting process, clearly showing 
that the comparison with real data does not work when one assumes
universal luminosity functions for different morphological types.
This view subscribes to the notion of passive evolution {\ie} an
initial distribution of galaxy types was imprinted in the early
Universe and very little has occurred by way of galaxy
transformational processes (such as major or minor galaxy mergers, ram
pressure stripping, and harrassment) to affect this makeup.  This view
falters when confronted with the high incidence of early-type galaxies
in the relatively shallow potential wells of groups.  

We argued in
\cite{miles04} that the persistent depletion of the
intermediate-luminosity galaxies in the optical luminosity functions
of X-ray dim groups shows that in the dynamically sluggish environment
of such groups (which have low velocity disperison), dynamical
friction would facilitate more rapid merging, thus depleting
intermediate-luminosity galaxies to form a few giant central galaxies,
resulting in the prominent dip seen in our LFs. In this work, we have
shown that this effect is also seen in the near-infrared LFs, but only
in the interior regions of the groups ($R<0.3\,R_{500}$), as in the
case of the optical LFs.  We also show that this effect vanishes when
one goes out to $R_{500}$, an effect we had not tested in our optical
CCD data due to lack of coverage. We thus show that the intermediate
depletion in the LFs is a real feature, and we argue that
it is due to the merging of
galaxies, rather than a bright-end enhancement caused by excess star
formation.

\section*{Acknowledgements}
Thanks to Trevor Ponman for interesting discussions and suggestions,
and, together with John Osmond, for their work on the X-ray properties
of the GEMS sample, and to Duncan Forbes and all our collaborators
in the multi-wavelength GEMS project.
This publication makes use of data products from
the Two Micron All Sky Survey, which is a joint project of the
University of Massachusetts and the Infrared Processing and Analysis
Center/California Institute of Technology, funded by the National
Aeronautics and Space Administration and the National Science
Foundation. This research has also made use of the NASA/IPAC
Extragalactic Database (NED) which is operated by the Jet Propulsion
Laboratory, California Institute of Technology, under contract with
the National Aeronautics and Space Administration.

\bsp

\label{lastpage}
\end{document}